\documentclass[11pt]{article}
\usepackage[a4paper,margin=25mm]{geometry}
\usepackage[T1]{fontenc}
\usepackage{lmodern}
\usepackage{amsmath,amssymb,braket,microtype}
\usepackage{authblk}
\title{A Hilbert-Space Analogy for Sexual Orientation and Gender Identity}
\author{Sofia Fatigoni}
\affil{California Institute of Technology, Pasadena, California, USA}
\date{}
\begin{document}
\maketitle
\begin{abstract}
We introduce a finite-dimensional state-space representation of bisexuality in Dirac notation, specify normalized amplitude weights, and analyze the limits of inference from one single observation. We extend the basis for discussing attraction across genders, then define a distinct state space for gender identity with an illustrative normalized time-dependent trajectory. The model is an analogy, not a physical theory or a measurement procedure for human sexuality.\end{abstract}
\section{Introduction}
Bisexuality can be modeled in a minimal representation by attraction to more than one gender. We specify a basis and normalized amplitudes, examine what can be inferred from an observed relationship, and extend the representation to additional categories. Gender identity is then considered in a separate state space. These constructions are formal analogies, not empirical quantum models of people.
\section{Sexual orientation: a two-state representation}
For a given person in an illustrative binary setting, let $\ket{S}$ represent attraction to someone of the same gender and $\ket{D}$ attraction to someone of the opposite gender. Treat these as orthonormal basis states, so $\braket{S|D}=0$ and $\braket{S|S}=\braket{D|D}=1$ \cite{dirac1939,dirac1958,vonneumann2018}. They are eigenstates of a formal category operator
\begin{equation}
\hat A=a_S\ket{S}\bra{S}+a_D\ket{D}\bra{D},\qquad
\hat A\ket{S}=a_S\ket{S},\quad \hat A\ket{D}=a_D\ket{D}.
\end{equation}
The eigenvalues $a_S$ and $a_D$ distinguish the two categories; the coefficients $\alpha$ and $\beta$ specify their weights in the state. The bisexual wavefunction in this simplified model is
\begin{equation}\label{eq:bi}
\boxed{\ket{\psi_{\rm bi}}=\alpha\ket{S}+\beta\ket{D}}.
\end{equation}
The coefficients are probability amplitudes. With real, nonnegative amplitudes we can write $\alpha=\sqrt{p_S}$ and $\beta=\sqrt{p_D}$; in general they may be complex, with $p_S=|\alpha|^2$ and $p_D=|\beta|^2$ \cite{dirac1958}. Under the formal measurement analogy, $|\alpha|^2$ and $|\beta|^2$ are the probabilities that a projection of the orientation state yields $\ket{S}$ or $\ket{D}$, respectively. These probabilities describe the model's projection outcomes; they do not, by themselves, predict the frequency of a person's relationships.

Crucially, the wavefunction must be normalized. Because the states are orthogonal,
\begin{equation}\label{eq:norm}
\boxed{\braket{\psi_{\rm bi}|\psi_{\rm bi}}=|\alpha|^2+|\beta|^2=p_S+p_D=1.}
\end{equation}
For example, $\sqrt{0.3}\ket{S}+\sqrt{0.7}\ket{D}$ is normalized since $0.3+0.7=1$. Equal weights are not required: bisexuality does not mean equal attraction to two genders. Within this two-state model, exclusively gay and exclusively straight orientations are limiting cases of Eq.~\eqref{eq:bi}: $\beta=0$ for the former and $\alpha=0$ for the latter.
\subsection{One observed eigenstate}
Consider the projectors $P_S=\ket{S}\bra{S}$ and $P_D=\ket{D}\bra{D}$, with $P_S+P_D=I$ \cite{sakurai2020,nielsen2010}. An observer who sees a bisexual woman dating a man registers one opposite-gender outcome; an observer who sees her dating a woman registers one same-gender outcome. A single outcome does not determine the coefficients of the full orientation vector. In particular, seeing $D$ does not imply $\beta=1$ or $\alpha=0$.
Projecting an observed relationship onto $\ket{S}$ or $\ket{D}$ classifies that relationship. It does not project the person's orientation into a single state: their full orientation remains represented by $\ket{\psi_{\rm bi}}$. A real relationship does not collapse a person's identity or erase other attractions. Calling someone straight or gay solely because of their current partner mistakes one observed relationship for the whole person \cite{nasem2022}.
\subsection{Extension to More Than Two Genders}
In the binary example, $\ket{D}$ represents attraction to the opposite gender. Introduce an orthogonal state $\ket{O}$ for attraction to genders outside that binary classification, including nonbinary genders. The extended state is
\begin{equation}
\boxed{\ket{\psi_3}=\alpha\ket{S}+\beta\ket{D}+\gamma\ket{O}},\qquad
\boxed{|\alpha|^2+|\beta|^2+|\gamma|^2=1.}
\end{equation}
Now $p_O=|\gamma|^2$ and $p_S+p_D+p_O=1$. Setting $\gamma=0$ gives the earlier two-state form. This three-state expansion illustrates attraction across the gender categories chosen for this model; it does not imply that attraction to nonbinary people is exclusive to pansexuality, as bisexual people may also experience such attraction \cite{nasem2022}. Nonbinary people are not one homogeneous third gender \cite{nasem2022}. Being transgender alone does not place a person in $\ket{O}$: a trans man is a man and a trans woman is a woman. Many pansexual people describe gender as irrelevant to their attraction \cite{nasem2022}. The basis vectors are coordinates for the analogy rather than fixed definitions of identity labels.
\section{Gender identity as a distinct state space}
Gender identity answers ``Who am I?''; sexual orientation answers ``To whom am I attracted?'' These are distinct questions. Neither determines the other: a person's gender identity does not specify the genders of people they can be attracted to \cite{nasem2022}. The orientation vector $\ket{\psi_{\rm bi}}$ defined above does not describe identity.

We can also use Dirac notation to represent gender identity. In the simplest illustrative basis, let $\ket{M}$ and $\ket{W}$ represent identification with ``man'' and ``woman,'' with $\braket{M|W}=0$. A nonbinary person who experiences aspects of both might choose to picture their identity as
\begin{equation}\label{eq:identity}
 \ket{\chi_{\rm NB}}=u\ket{M}+v\ket{W},\qquad |u|^2+|v|^2=1.
\end{equation}
This depicts one possible nonbinary experience. Nonbinary identity does not require being partly man and partly woman; a person may identify with neither or describe their gender in entirely different terms \cite{nasem2022}. More basis states, or a different basis, would be needed to represent those descriptions.

\subsection{Time-dependent representation}
Time dependence can represent a particular account of changing self-understanding. For example, someone who once identified with $\ket{M}$ and increasingly identifies with $\ket{W}$ might, if this fits their own account, choose the illustrative path
\begin{equation}\label{eq:transition}
 \ket{\chi(t)}=e^{-t/\tau}\ket{M}
 +\sqrt{1-e^{-2t/\tau}}\ket{W},\qquad t\geq0,\quad\tau>0.
\end{equation}
Here $\tau$ is a freely chosen timescale for this example. The factor $e^{-t/\tau}$ multiplies one amplitude, so its squared weight is $e^{-2t/\tau}$; the second amplitude keeps the state normalized at every time:
\begin{equation}
 \braket{\chi(t)|\chi(t)}=e^{-2t/\tau}+(1-e^{-2t/\tau})=1.
\end{equation}
This curve could symbolize one person's journey of recognizing or expressing their gender. It is not a general model of being transgender: many trans people have always known their gender, do not experience a fading identity, or identify outside this binary basis. There is no universal transition timescale. A trans woman is a woman and a trans man is a man; ``trans'' is not a third gender eigenstate.

Formally, identity and orientation may be written in separate spaces, $\mathcal H_{\rm identity}$ and $\mathcal H_{\rm orientation}$. If we combine the bookkeeping, a simple independent example is $\ket{\chi}\otimes\ket{\psi}$: specifying either factor alone does not fix the other \cite{nielsen2010}. Real lives need not obey a mathematical independence assumption; ``uncoupled'' here means that neither label logically determines the other.

\section{Domain and limitations}
The amplitudes are formal coordinates, not empirical frequencies of relationships. Partner choice involves opportunity, preference, reciprocity, and circumstance. An observer does not perform a controlled projective measurement of a person's orientation or identity; the mathematics entails no physical collapse, interference, or quantum dynamics. The formalism establishes only an inference constraint: one observed relationship does not specify a person's full orientation.
\section{Conclusion}
The normalized two-state representation illustrates bisexuality and the information omitted by a single observed pairing. A refined basis extends the classification to further attraction categories without overlap. A distinct identity space supports a separate, normalized time-dependent example. The validity of a person's identity and orientation rests with their self-description, not with an externally assigned state vector.

\bibliographystyle{unsrt}
\bibliography{references}

\end{document}